\documentclass[journal,twoside]{IEEEtran}
\usepackage[OT1]{fontenc}

\usepackage{amsmath}
\usepackage{amsthm}
\usepackage{amssymb}
\usepackage[linesnumbered,lined,ruled,vlined]{algorithm2e}
\usepackage{bm}
\usepackage{booktabs}
\usepackage{cite}
\usepackage{cases}
\usepackage[usenames, dvipsnames]{color}
\usepackage{colortbl}
\usepackage{float}
\usepackage[colorlinks, urlcolor=blue, linkcolor=black, anchorcolor=black, citecolor=black]{hyperref}
\usepackage{url}
\usepackage{graphicx}
\usepackage{subfigure}
\usepackage{lipsum}
\usepackage{makecell} 
\usepackage{multirow}
\usepackage{microtype} 
\usepackage{mathtools}
\usepackage{subfigure}
\usepackage{tabularx}
\usepackage{threeparttable}  
\usepackage{xfrac}
\usepackage{balance}
\usepackage{xcolor}
\usepackage{stfloats} 
\usepackage{diagbox} 
\usepackage{pgfplots} 

\definecolor{mygray}{gray}{.9}

\theoremstyle{definition}

\definecolor{mygreen}{HTML}{00B050}

\begin{document}
	
	\title{Spatiotemporal 2-D Channel Coding for Very Low Latency Reliable MIMO Transmission}
	\author{Xiaohu You,~\IEEEmembership{Fellow,~IEEE},
            Chuan Zhang,~\IEEEmembership{Senior Member,~IEEE},
            Bin Sheng,\\
            Yongming Huang,~\IEEEmembership{Senior Member,~IEEE},
            Chen Ji,
            Yifei Shen,
            Wenyue Zhou,
            and Jian Liu
\thanks{X. You, C. Zhang, B. Sheng, Y. Huang, Y. Shen, W. Zhou, and J. Liu are with the National Mobile Communications Research Laboratory, Southeast University, Nanjing 211189, China, and also with the Purple Mountain Laboratories, Nanjing 211189, China. email: \{xhyu,chzhang\}@seu.edu.cn. \emph{(Corresponding authors: Xiaohu You and Chuan Zhang.)}}
\thanks{C. Ji is with the School of Information Science and Technology, Nantong University, Nantong 226019, China.}
	}
	
	\markboth{~~}%
	{X. You \MakeLowercase{\textit{et al.}}: Spatiotemporal 2-D Channel Coding for Very Low Latency Reliable MIMO Transmission}
	

	\maketitle
	
	\begin{abstract}		
	To fully support vertical industries, 5G and its corresponding channel coding are expected to meet requirements of different applications. However, for applications of 5G and beyond 5G (B5G) such as URLLC, the transmission latency is required to be much shorter than that in eMBB. Therefore, the resulting channel code length reduces drastically. In this case, the traditional 1-D channel coding suffers a lot from the performance degradation and fails to deliver strong reliability with very low latency. To remove this bottleneck, new channel coding scheme beyond the existing 1-D one is in urgent need. By making full use of the spacial freedom of massive MIMO systems, this paper devotes itself in proposing a spatiotemporal 2-D channel coding for very low latency reliable transmission.
For a very short time-domain code length $N^{\text{time}}=16$, $64 \times 128$ MIMO system employing the proposed spatiotemporal 2-D polar channel coding scheme successfully shows more than $3$\,dB performance gain at $\text{FER}=10^{-3}$, compared to the 1-D time-domain channel coding.
It is noted that the proposed coding scheme is suitable for different channel codes and enjoys high flexibility to adapt to difference scenarios. By appropriately selecting the code rate, code length, and the number of codewords in the time and space domains, the proposed coding scheme can achieve a good trade-off between the transmission latency and reliability.
	\end{abstract}
	

	\IEEEpeerreviewmaketitle

	\section{Introduction}

\IEEEPARstart{W}{ith} its rapid development, wireless communications now ranks as a key driver of economic and social progress. Officially commercialized in the year of 2019 \cite{3GppCommercial}, the fifth generation network (5G) of cellular mobile communications has become the leading part of the ``new infrastructure'', and greatly transforms vertical industries. To fulfill the demands of different vertical applications, 5G is sliced into three use cases: enhanced mobile broadband (eMBB), ultra-reliable low-latency communications (URLLC), and massive machine type communications (mMTC) \cite{5Gservicerequire, 5Gservicerequire2}. As one of the most crucial parts of baseband processing, channel coding for 5G now faces challenging requirements on flexibility, decoding performance, decoding latency, implementation complexity, and so on. One well-known example is the eMBB case adopts two kind of channel codes at the same time: low-density parity-check (LDPC) codes used for data channels, and polar codes used for control channels \cite{5Gphysicallayer, 5Gchannelcode}.

However, in order to achieve the goals set for 5G, such hybrid channel coding setup is far from enough. Further enabling channel coding schemes are expected. One existing challenge is the implementation of very low latency reliable multiple-input multiple-output (MIMO) transmission for URLLC. In the eMBB case, both channel codes (LDPC codes and polar codes) are coded in one dimension, time domain. Admittedly, this 1-D coding is acceptable for eMBB, since the corresponding package length is sufficiently long for both LDPC codes and polar codes to deliver satisfactory decoding performance \cite{5Gservicerequire2}. Unfortunately, for URLLC, in order to lower the latency, the package length is much reduced \cite{5Gservicerequire2, 5Gurllc}. Consequently, the decoding performance heavily degrades and the transmission reliability is no longer guaranteed. Therefore, new coding schemes which can flexibly balance both transmission reliability and decoding latency is highly required.

On the other hand, it is noticed that massive MIMO technology is another enabling technique for 5G besides channel coding \cite{5GNRtechnology}. It brings considerable advantages in both spectral efficiency (SE) and energy efficiency (EE). For channel codes, the massiveness of antennas offers another dimension, space domain for encoding \cite{you2020shannon}. This enables us to propose a spatiotemporal 2-D channel coding to implement very low latency reliable data transmission for 5G. The proposed scheme introduce the space-domain coding over the current time-domain coding. With very short time-domain code length, it can greatly improve the decoding performance. This scheme cannot only meet URLLC's requirements on latency, reliability, and transmission rate, but also enjoy perfect flexibility for applications in various scenarios.

The contributions of this paper are as follows.

\begin{itemize}
  \item Making use of both the time domain and space domain of massive MIMO systems, this paper proposes a spatiotemporal 2-D channel coding scheme which can well balance transmission latency and reliability.
  \item A configuration method by leveraging the layer mapping, code length, code rate to flexibly meet requirements of different applications scenarios including eMBB and URLLC is proposed.
  \item The proposed scheme is general, regular, and implementation-friendly. It can be applied to any existing MIMO systems without touching the other modules and has wide application prospect.
\end{itemize}

The remainder of this paper is organized as follows. Section~\ref{sec:preliminaries} introduces the multi-antenna baseband transmitting systems, where the layer mapping and time-domain 1-D coding are described in detail. Section~\ref{sec:2DCCS} proposes the spatiotemporal 2-D channel coding scheme with the corresponding baseband system and latency analysis. Numerical results are provided in Section~\ref{sec:results}. The advantages of the proposed coding scheme are discussed in Section~\ref{sec:discussions} and Section~\ref{sec:conclusion} concludes this paper.

    \section{Preliminaries}\label{sec:preliminaries}
In this section, a brief introduction of the traditional multi-antenna baseband transmitting system, the layer mapping, and the time-domain 1-D channel coding scheme is given. Since this paper is on the channel coding scheme, we mainly focus on the corresponding transmitting system. Readers who would like to see more details of the entire system including the receiving part, can refer to Fig.~1 of \cite{zhang2017advanced}.

\subsection{Multi-Antenna Baseband Transmitting Systems}\label{sec:preliminaries:1}
\begin{figure*}[t]
  \centering
  \includegraphics[width=0.95\linewidth]{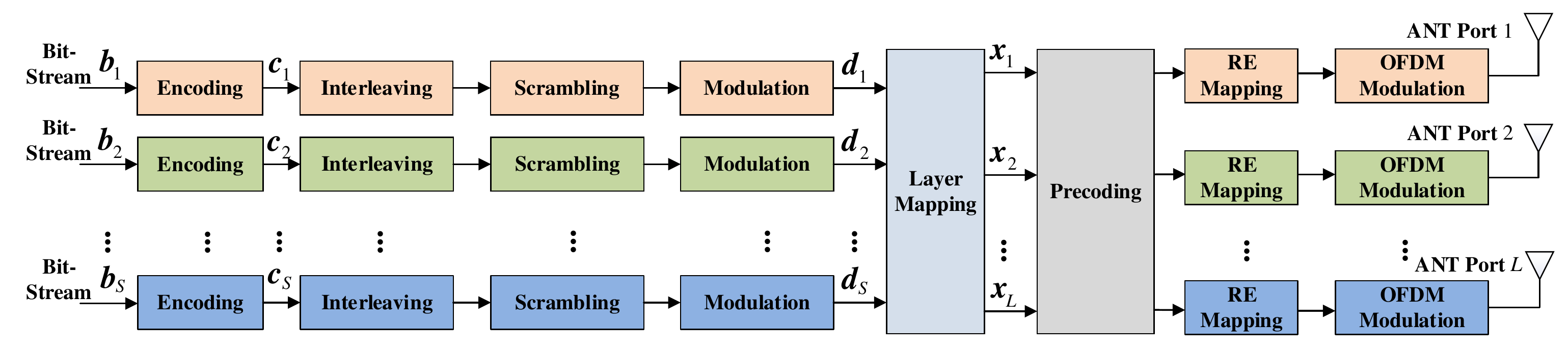}
  \caption{The block diagram of the multi-antenna baseband transmitting system.}
  \label{fig:5G_baseband}
\end{figure*}

In the physical (PHY) layer of wireless systems, the transmission block (TB) from the media access control (MAC) layer is transferred into multiple orthogonal frequency division multiplexing (OFDM) symbols and then transmitted via multiple antenna ports (ANT ports). The details of a multi-antenna baseband transmitting system are illustrated in Fig.~\ref{fig:5G_baseband}.

The input TB is first divided into several bit-streams $\{\bm{b}_s\}$, where $1\leq s\leq S$ and $S$ is the cardinality of $\{\bm{b}_s\}$. Each bit-stream $\bm{b}_s$ is of length $K_s$, and will be encoded into a codeword $\bm{c}_s$ of length $N_s$ and rate $R_s=K_s/N_s$. After the operations of interleaving, scrambling, and modulation, $\bm{c}_s$ is mapped to a complex-valued symbol sequence $\bm{d}_s$. For different modulation schemes, each symbol corresponds to different number of bits. For instance, with $2^q$-quadrature amplitude modulation (QAM), each symbol of $\bm{d}_s$ is mapped from $q$ continuous bits.



For air interface transmission, those symbols are mapped to ANT ports by layer mapping, precoding, resource element (RE) mapping, and OFDM modulation. The detailed steps are as follows.


By layer mapping, $\bm{d}_s$ is allocated into several layers out of totally $L$ layers. Without loss of generality, in this paper it is assumed that the number of layers is same as the number of ANT ports and the number of transmitting antennas (Fig.~\ref{fig:5G_baseband}).
In Layer $l$ ($1\leq l\leq L$), the symbol sequence is denoted as $\bm{x}_l$, which is further precoded with the precoding matrix. The precoded symbols are first scaled with the amplitude factor regarding the transmitting power, and then mapped to the REs. Finally, these scaled symbols are modulated to orthogonal subcarriers, and transmitted via ANT ports in the synthetic OFDM symbols \cite{Jonna2013rank}. In the aforementioned steps, the layer mapping is of good relevance of the proposed channel coding scheme and is detailed in the following sub-section.


\subsection{Layer Mapping}



The layer mapping scheme is proposed due to the inherently differentiated demands of wireless communications.
In practice, the wireless systems will involve different users/tasks, and different users/tasks need to meet different requirements. Therefore, layer mapping enables us to apply different schemes such as channel coding schemes (i.e., code type, code rate, code length, and so on) in different layers. In real applications, there are two mainstream layer mapping schemes, namely 1) the folded layer mapping and 2) the parallel layer mapping.

\subsubsection{Folded Layer Mapping}
\begin{figure}[htbp]
  \centering
  \includegraphics[width=\linewidth]{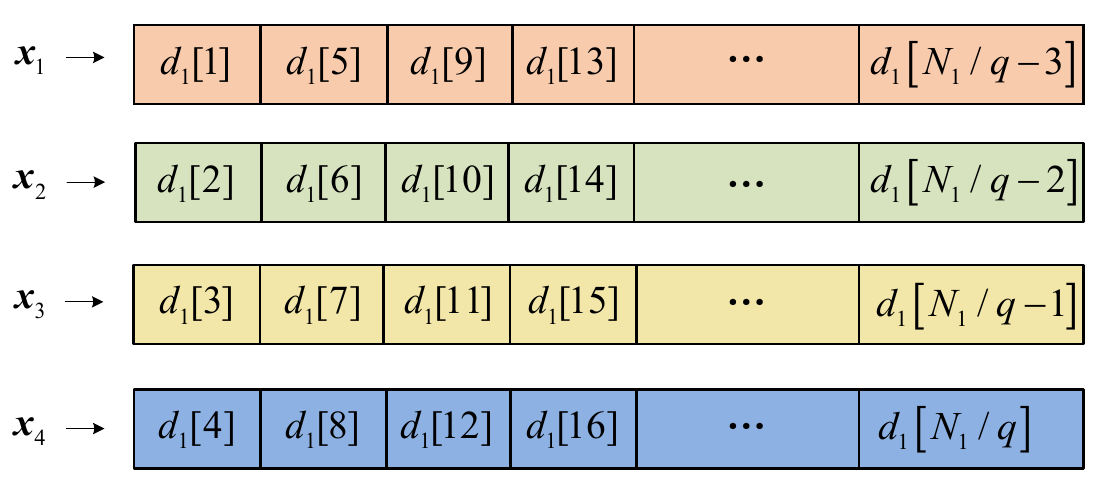}\\
  \caption{The illustration of the folded layer mapping with $L=4$ and $S=1$.}\label{fig:mapB}
\end{figure}

Since the implementation cost increases linearly with the number of bit-streams ($S$), in traditional applications which are sensitive to complexity, there is sometimes a constraint on $S$. For example, the current 5G NR sets the maximum number of bit-streams as $2$ \cite{5Gphysical}. Therefore, it holds that $S<L$ and the layer mapping needs to be done by folding one bit-stream into multiple layers. Fig.~\ref{fig:mapB} shows an example of the folded layer mapping with $L=4$ and $S=1$, which is compatible with the current 5G NR.

\subsubsection{Parallel Layer Mapping}
\begin{figure}[htbp]
  \centering
  \includegraphics[width=\linewidth]{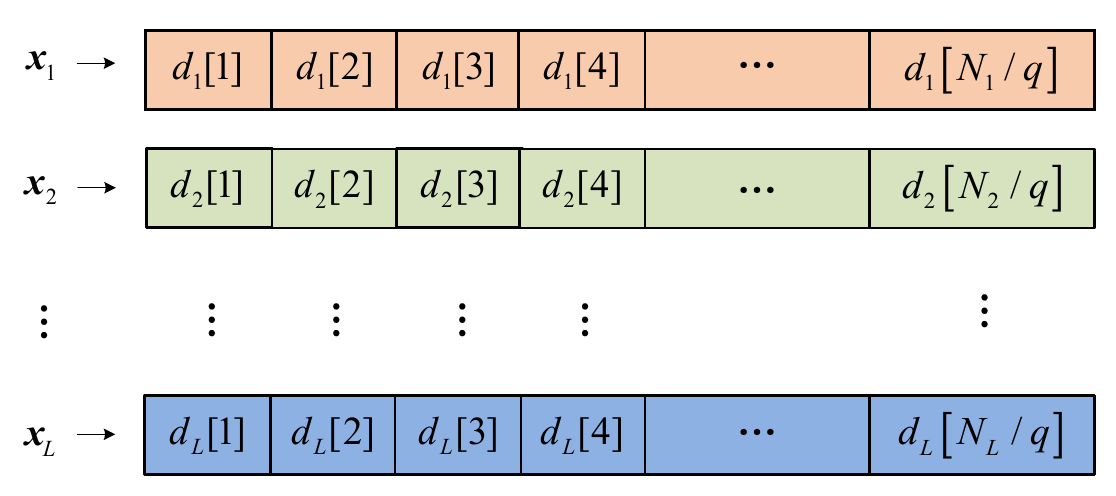}\\
  \caption{The illustration of the parallel layer mapping, where $S=L$.}\label{fig:mapA}
\end{figure}

However, with the progressive evolution of 5G NR, implementation complexity is no longer the key bottleneck. And the existing folded layer mapping shows disadvantages such as high latency and low flexibility. Though straightforward but efficient, the parallel layer mapping becomes favorable. In this scheme, the number of layers ($L$) is set the same as the number of bit-streams ($S$). More specifically, for the parallel layer mapping, we have $S=L$ and $\bm{x}_l=\bm{d}_s$ for $\forall~l=s$. Therefore, all bit-streams can be encoded independently and in parallel, which reduces latency and improves both flexibility and coding efficiency.
Illustration of the parallel layer mapping is given by Fig.~\ref{fig:mapA}.



\subsection{Time-Domain 1-D Channel Coding}
According to Section \ref{sec:preliminaries:1}, it can be seen that for either the parallel layer mapping or the folded layer mapping, the channel coding is done in 1-D time domain. Taking the parallel layer mapping as an example, the time-domain codewords are of length $N_s$ and rate $R_s=K_s/N_s$.

\begin{figure}[htbp]
  \centering
  \begin{tikzpicture}
    \definecolor{myblued}{RGB}{0,114,189}
    \definecolor{myred}{RGB}{217,83,25}
    \definecolor{myyellow}{RGB}{237,137,32}
    \definecolor{mypurple}{RGB}{126,47,142}
    \definecolor{myblues}{RGB}{77,190,238}
    \definecolor{mygreen}{RGB}{32,134,48}
    \definecolor{mypink}{RGB}{255,62,150}
      \pgfplotsset{
        label style = {font=\fontsize{9pt}{7.2}\selectfont},
        tick label style = {font=\fontsize{7pt}{7.2}\selectfont}
      }

    \begin{axis}[
        scale = 1,
        ymode=log,
        xmin=10.0,xmax=18.0,
        ymin=3.0E-06,ymax=0.8,
        xlabel={$E_b/N_0$ [dB]}, xlabel style={yshift=0.1em},
        ylabel={FER}, ylabel style={yshift=-0.75em},
        xtick={10,11,12,13,14,15,16,17,18},
        xticklabels={10,11,12,13,14,15,16,17,18},
        grid=both,
        ymajorgrids=true,
        xmajorgrids=true,
        grid style=dashed,
        width=0.9\linewidth,
        thick,
        mark size=1,
        legend style={
          nodes={scale=0.5, transform shape},
          anchor={center},
          cells={anchor=west},
          column sep= 1mm,
          row sep= -0.25mm,
          font=\fontsize{8pt}{7.2}\selectfont,
        },
        legend columns=1,
        legend pos=south west,
    ]

    \addplot[
        color=myyellow,
        mark=square*,
        densely dotted,
        every mark/.append style={solid},
        line width=0.25mm,
        mark size=1.9,
        fill opacity=0,
    ]
    table {
     10     0.699300699300699	
     11     0.436681222707424	
     12     0.251256281407035	
     13     0.174216027874564	
     14     0.106044538706257	    
     15     0.0493339911198816	     
     16     0.0279720279720280	    
     17     0.0159159637116027	    
     18     0.00657462195923734	    
     19     0.00308194902456313	    
     20     0.00114005586273727
    };
    \addlegendentry{Time-domain 1-D, $N_s=16$, $R_s=1/4$}

    \addplot[
        color=myblues,
        mark=diamond*,
        densely dotted,
        every mark/.append style={solid},
        line width=0.25mm,
        mark size=1.9,
        fill opacity=0,
    ]
    table {
     10     0.492610837438424	
     11     0.238663484486874	
     12     0.130548302872063	
     13     0.0652741514360313	
     14     0.0381533765738268	    
     15     0.0176772140710624	     
     16     0.00905305087814593	
     17     0.00473596968979399	    
     18     0.00241027742293138	    
     19     0.000792148226776194	
     20     0.000248765501200294
    };
    \addlegendentry{Time-domain 1-D, $N_s=32$, $R_s=1/4$}

    \addplot[
        color=myred,
        mark=triangle*,
        densely dotted,
        every mark/.append style={solid},
       line width=0.25mm,
        mark size=1.9,
        fill opacity=0,
    ]
    table {
     10     0.170068027210884	
     11     0.063011972274732	
     12     0.0226500566251416	
     13     0.00828706389326262	
     14     0.00348298561526941	    
     15     0.00138969954695795	     
     16     0.000373254103928873	
     17     0.000133505109908082   	
     18     3.40000000000000e-05	
     19     1.40000000000000e-05
    };
    \addlegendentry{Time-domain 1-D, $N_s=64$, $R_s=1/4$}

    \addplot[
        color=mypurple,
        mark=*,
        densely dotted,
        every mark/.append style={solid},
        line width=0.25mm,
        mark size=1.6,
        fill opacity=0,
    ]
    table {
     10     0.072780203784570	
     11     0.016246953696182	
     12     0.005243013684265	
     13     0.0018285881470916	
     14     0.0005808752628460  	
     15     0.000144735394751316	 
     16     5.20000000000000e-05	
     17     1.30000000000000e-05	
     18     4.00000000000000e-06
    };
    \addlegendentry{Time-domain 1-D, $N_s=128$, $R_s=1/4$}
    \end{axis}
    \end{tikzpicture}
  \caption{FER performance of time-domain 1-D channel coding with decreasing code lengths. 
  }\label{fig:results01}
\end{figure}
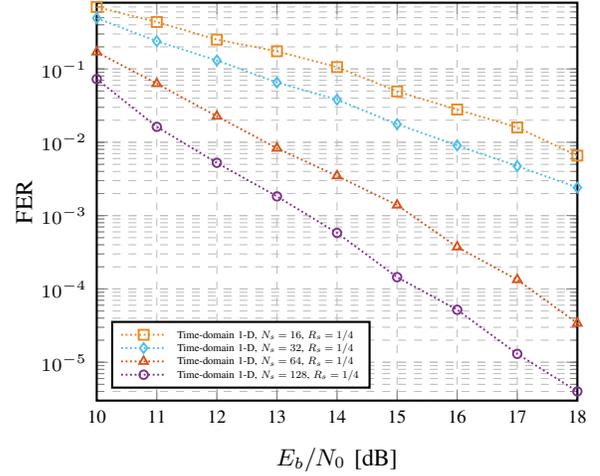

Admittedly, the current time-domain 1-D channel coding scheme can sufficiently meet the requirements posed by eMBB. However, for other applications such as URLLC, which requires very low latency and ultra-reliable transmission, the time-domain channel coding scheme no long serves. For URLLC, in order to meet the latency requirement, the package length or codeword length ($N_s$) should be much shorter than that in eMBB, which will inevitably reduce the transmission reliability. According to Fig.~\ref{fig:results01}, even for the parallel layer mapping, the frame error rate (FER) performance of 1-D time-domain coding degrades severely as $N_s$ reduces. In this plot, polar code using Tal-Vardy construction \cite{talconstruct:13} is chosen as the channel code and the MIMO antenna configuration is $N_t \times N_r = 64 \times 128$. The modulation scheme is $16$-QAM. The polar decoder is the successive cancellation (SC) decoder \cite{arikan2009channel}. The MIMO detector is the minimum mean square error (MMSE) detector \cite{MMSE:02, MMSE:imp, MMSE:softoutput}. Without loss of generality, the above configuration is adopted as the running setup for the following simulations, unless specified additionally.

Therefore, new channel coding schemes which can offer very low latency and ultra-reliable transmission is in urgent need. To this end, the rest of this paper proposes a spatiotemporal 2-D channel coding scheme.

    \section{Proposed Spatiotemporal 2-D Channel Coding}\label{sec:2DCCS}
In this section, the proposed spatiotemporal 2-D channel coding scheme is introduced with details. Its flexibility and latency analysis are also given.

\subsection{Multi-Antenna Transmitting Systems with 2-D Coding}\label{sec:2DCCS-motivation}


To address the contradiction between latency and reliability of 1-D coding, we need to revisit the multi-antenna transmitting systems in Fig.~\ref{fig:5G_baseband}. For massive MIMO systems, the number of ANT ports ($L$) is relatively large and provides us another dimension for operation, i.e. the space domain. This inspires us to improve the reliability by making use of the already existing antenna massiveness. The proposed scheme is named spatiotemporal 2-D channel coding scheme, which introduces the space-domain coding over the current time-domain coding. This scheme cannot only meet URLLC's requirements on latency, reliability, and transmission rate, but also enjoy perfect flexibility for applications in various scenarios.



\begin{figure*}[htbp]
  \centering
  \includegraphics[width=1\linewidth]{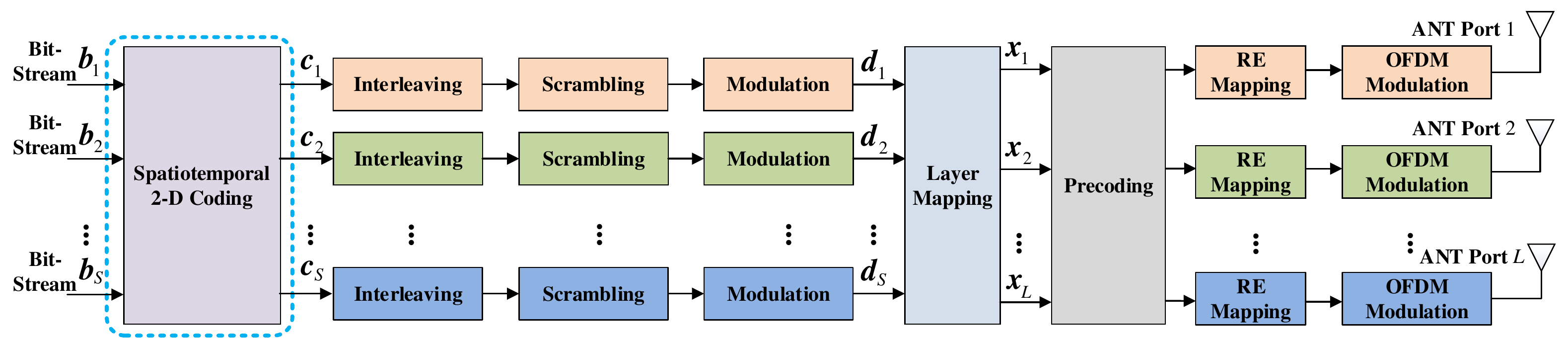}\\
  \caption{The block diagram of the multi-antenna baseband transmitting system with the spatiotemporal 2-D coding.}\label{fig:sysB}
\end{figure*}

Fig.~\ref{fig:sysB} shows the details of a multi-antenna baseband transmitting system with 2-D channel coding. Compared to the system in Fig.~\ref{fig:5G_baseband}, the \verb|Encoding| modules have been replaced by a \verb|Spatiotemporal 2-D Coding| module, whereas the other modules stay unchanged. In general, the proposed spatiotemporal 2-D coding can be directly applied to any 1-D coded systems without touching the other modules except the coding module. Therefore, this scheme expects wide applications. Similar to the 1-D coded system in Fig.~\ref{fig:5G_baseband}, the \verb|Layer Mapping| module distributes the complex-valued symbols in $L$ layers. In each layer, the number of symbols is $M_{\text{sym}}$. In practice, $M_{\text{sym}}$ is determined by the input length of the system and bounded by the number of subcarriers, which is very small for URLLC.

Before giving details of the spatiotemporal 2-D channel coding scheme, the concept of the codeword trellis is introduced in the next subsection.


\subsection{Codeword Trellis}



\begin{figure}[htbp]
  \centering
  \includegraphics[width=0.85\linewidth]{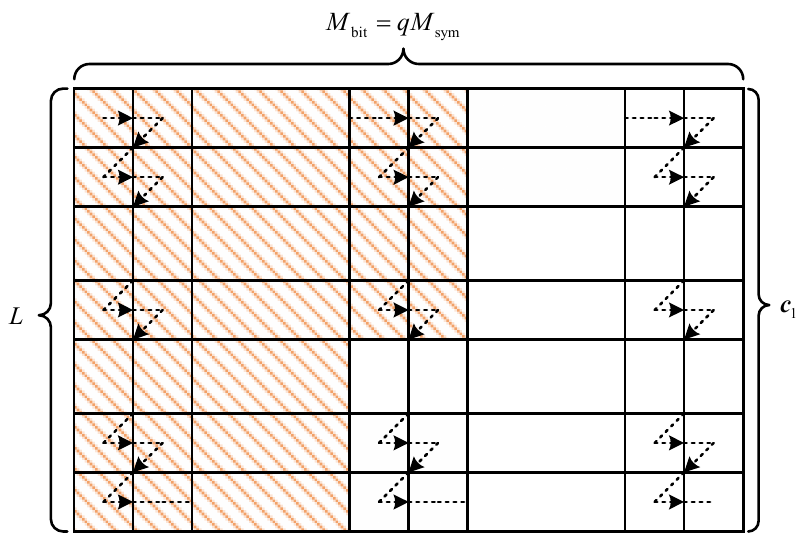}\\
  \caption{The codeword trellis of the folded layer mapping with $S=1$.}\label{fig:trellisB}
\end{figure}

\begin{figure}[htbp]
  \centering
  \includegraphics[width=0.85\linewidth]{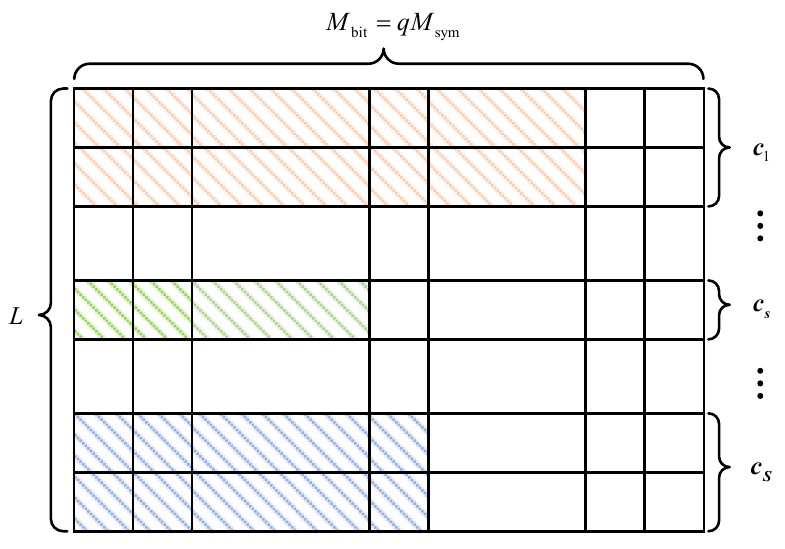}\\
  \caption{The codeword trellis of the parallel layer mapping with flexibility, when $S=L$ it corresponds to Fig.~\ref{fig:mapA}.}\label{fig:trellisA}
\end{figure}

In each layer, $M_{\text{sym}}$ complex-valued symbols are mapped from $M_{\text{bit}}=qM_{\text{sym}}$ coded bits. Therefore, the codewords before modulation form a (bit-level) \emph{codeword trellis}, whose width and height are $M_{\text{bit}}$ and $L$, respectively.


Assume the area of the TB is fixed. Given the transmission latency $(M_{\text{bit}})$ and the number of layers $(L)$,
different coding schemes result in different distributions of information bits in the codeword trellis. And therefore, different decoding performances (transmission reliability) are obtained. The key task of us is to find out the distribution, which can well balance transmission latency and reliability. The codeword trellis is a generalized representation of the illustrations in both Fig.s~\ref{fig:mapB} and \ref{fig:mapA}. For example, the codeword trellis of the folded layer mapping is in Fig.~\ref{fig:trellisB}, where $4$-QAM is employed and information bits are shaded. And the codeword trellis of the parallel layer mapping is shown in Fig.~\ref{fig:trellisA}. The rest part of this paper employs the codeword trellis extensively.

\subsection{Spatiotemporal 2-D Channel Coding}\label{subsec:flexible2DTS}
As discussed previously, for URLLC the transmission latency reduces and $S$ increases. The resulting degraded reliability is now compensated by the space-domain coding. Herein, we introduce the spatiotemporal 2-D channel coding scheme with the help of codeword trellis. We first consider the time-space mode: the time-domain coding is performed first, followed by the space-domain coding. The number of codewords in time domain is $S$ and the number of codewords in space domain is $T$.



\begin{figure}[t]
  \centering
  \includegraphics[width=0.95\linewidth]{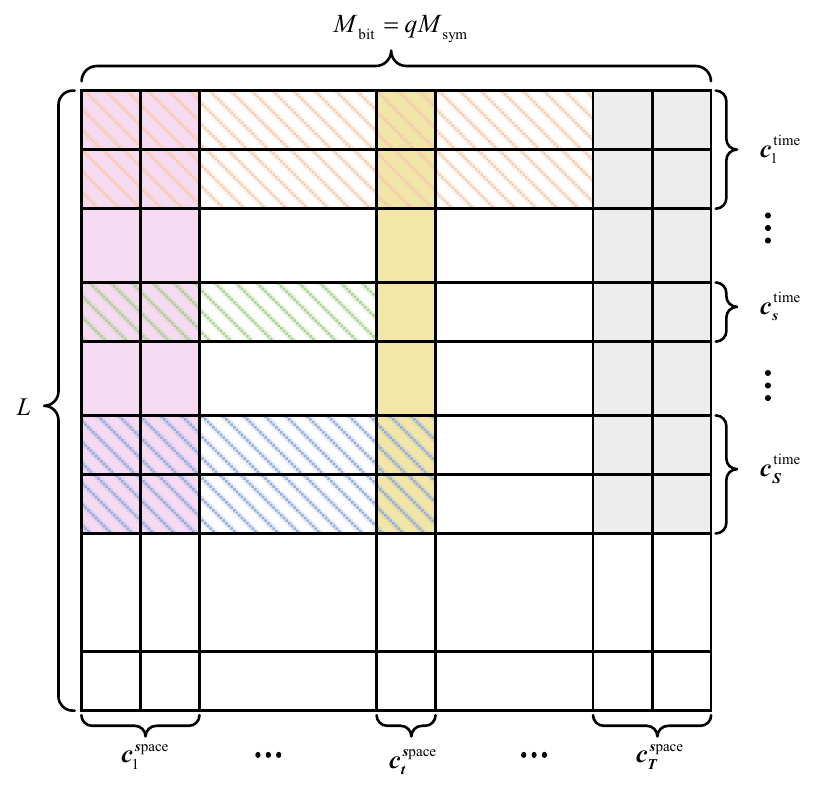}\\
  \caption{The codeword trellis of the spatiotemporal 2-D channel coding.}\label{fig:trellisFlexible}
\end{figure}

The codeword trellis of the time-space mode is shown in Fig.~\ref{fig:trellisFlexible}. In the time domain, the $S$ bit-streams $\{\bm{b}_s\}$ are first encoded to $S$ codewords $\{\bm{c}^{\text{time}}_s\}$, where $1\leq s\leq S$. For each codeword $\bm{c}^{\text{time}}_s$, the length and rate are $N^{\text{time}}_s$ and $R^{\text{time}}_s$, respectively. It is noted that the rate-matching is required to make sure $N^{\text{time}}_s$ is a multiple of $M_{\text{bit}}$, so that the codeword $\bm{c}^{\text{time}}_s$ is mapped to $N^{\text{time}}_s/M_{\text{bit}}$ layers. Similarly, in the space domain, there are $T$ codewords $\{\bm{c}^{\text{space}}_t\}$, whose lengths $N^{\text{space}}_t$ and rates $R^{\text{space}}_t$ can be different ($1\leq t\leq T$). The rate-matching is also required in the space domain.

The code type, code rate, and code length of each codeword in either time or space domain can be flexibly determined upon the requirements. At the receiving system, the log-likelihood ratios (LLRs) of each $\bm{c}_{t}^{\text{space}}$ are obtained first, by $T$ soft-output decoders. After the space-domain decoding, the LLRs of each $\bm{c}_{s}^{\text{time}}$ are obtained and $S$ decoders are activated to recover the information bits.

\noindent\textbf{{Example:}}
For better understanding of the proposed scheme, an example with $M_{\text{bit}}=16$ and $L=64$ is given here. Suppose there are $256$ information bits $\{a_1\ldots a_{256}\}$ in the TB, which are lined in parallel to form $32$ bit-streams. Assume three code rates are required according to the feedback from the receiver, which are $1/4$ for $\{\mathbf{b}_1 \ldots \mathbf{b}_{10}\}$, $1/2$ for $\{\mathbf{b}_{11}\ldots\mathbf{b}_{22}\}$, and $3/4$ for $\{\mathbf{b}_{23}\ldots\mathbf{b}_{32}\}$.
The bit-streams are explicitly given as follows:
\begin{equation}
\left\{
\begin{aligned}
&\bm{b}_1 =[a_1,a_2,a_3,a_4],\\
& \ \ \ \vdots\\
&\bm{b}_{10}=[a_{37},a_{38},a_{39},a_{40}],\\
&\bm{b}_{11}=[a_{41},a_{42},a_{43},a_{44},a_{45},a_{46},a_{47},a_{48}],\\
&\ \ \ \vdots\\
&\bm{b}_{22}=[a_{129},a_{130},a_{131},a_{132},a_{133},a_{134},a_{135},a_{136}],\\
&\bm{b}_{23}=[a_{137},a_{138},\dots,a_{148}],\\
&\ \ \ \vdots\\
&\bm{b}_{32}=[a_{244},a_{245},\dots,a_{256}].\\
\end{aligned}
\right.
\label{eq:TS_2_03}
\end{equation}
Based on the specified coding scheme, $32$ codewords are generated with the same code length $N^\text{time}=16$.
\begin{equation}
\resizebox{0.91\linewidth}{!}{$
\bm{c}^{\text{time}}\!\!=\!\!\left[ \begin{array}{c}
	\bm{c}^{\text{time}}_1\\
	\bm{c}^{\text{time}}_2\\
	\vdots\\
	\bm{c}^{\text{time}}_{32}\\
\end{array} \right] \!\!=\!\!\left[ \begin{matrix}
	\!\bm{c}^{\text{time}}_1[1]\!&		\!\bm{c}^{\text{time}}_1[2]\!&		\cdots&		\!\bm{c}^{\text{time}}_1[16]\!\\
	\!\bm{c}^{\text{time}}_2[1]\!&		\!\bm{c}^{\text{time}}_2[2]\!&		\cdots&		\!\bm{c}^{\text{time}}_2[16]\!\\
	\vdots&		\vdots&		\ddots&		\vdots\\
	\!\bm{c}^{\text{time}}_{32}[1]\!&		\!\bm{c}^{\text{time}}_{32}[2]\!&		\cdots&		\!\bm{c}^{\text{time}}_{32}[16]\!\\
\end{matrix} \right].$}
\label{eq:TS_4}
\end{equation}
Further, each column is encoded in the space domain with the same code rate $R^{\text{space}}=1/2$,
\begin{equation}
\!\bm{c}^{\text{space}}_t\!=\![\bm{c}^{\text{time}}_1[t],\dots,\bm{c}^{\text{time}}_{32}[t]]\cdot\mathbf{G}^{\text{space}},~~1\leq t\leq 16,
\end{equation}
where $\mathbf{G}^{\text{space}}$ is the generator matrix of the codeword in the space domain.



It is noted that we can also perform the space-domain coding first, then the time-domain coding. This mode is named space-time mode. Consequently, the receiver first takes care of the time-domain decoding, then the space-domain decoding.

\subsection{Latency Analysis}\label{subsec:latency}

According to the above analysis, different coding schemes over the same codeword trellis have the same transmission latency but different decoding latencies. In pipelined transmission, the transmission process and decoding process are overlapped. Therefore, the resulting latency is calculated in a synthetic manner with the following three assumptions:

1) The decoding cannot start unless the LLRs of a complete codeword trellis are received.

2) All decoders in the same domain can work in parallel.

3) The latency of decoding a codeword $\bm{c}_s$ is proportional to the code length, as $\mathcal{D}(\bm{c}_s)=\gamma \cdot N_s$ (SC polar decoder is assumed).

Therefore, the decoding latency of the proposed spatiotemporal 2-D channel coding scheme is:
\begin{equation}\label{eq:decodingLatency}
\mathcal{D}=\gamma\cdot\left(\max\limits_{1\leq s\leq S}\{N_s^{\text{time}}\}+\max\limits_{1\leq t\leq T}\{N_t^{\text{space}}\}\right).
\end{equation}

The lowest latency $\mathcal{D}_{\min}=\gamma\cdot(M_{\text{bit}}+L)$ is achieved when each codeword in the time/space domain takes up exactly one row/column of the codeword trellis. The decoding latency increases as $S$ or $T$ decreases.

The coding schemes in Fig.s~\ref{fig:trellisB} and \ref{fig:trellisA} are two special cases of the proposed spatiotemporal 2-D channel coding scheme. In Fig.~\ref{fig:trellisB}, we have $S=L$ and $R^{\text{space}}_t=1$. Its decoding latency is $\mathcal{D}=\gamma\cdot(M_{\text{bit}})$, which is low but with a poor decoding performance as shown in Fig.~\ref{fig:results01}. In Fig.~\ref{fig:trellisA}, no space-domain coding is involved and $S=1$. Its decoding latency is $\mathcal{D}=\gamma\cdot(L\cdot M_{\text{bit}})$, which is very long. Therefore, spatiotemporal 2-D channel coding scheme considering both time and space domain coding is a good choice for balancing latency and reliability. 

    \section{Numerical Results}\label{sec:results}
In this section, numerical results are provided to demonstrate the advantages of the proposed spatiotemporal 2-D channel coding scheme. The simulation setup keeps the same as that of Fig.~\ref{fig:results01} unless specified additionally. In order to achieve the lowest latency $\mathcal{D}_{\min}=\gamma\cdot(M_{\text{bit}}+L)$, it is assumed that each codeword in the time/space domain takes up exactly one row/column of the codeword trellis. The corresponding time/space-domain code length and code rate are $N^{\text{time}}$/$N^{\text{space}}$ and $R^{\text{time}}$/$R^{\text{space}}$, respectively.


\subsection{Spatiotemporal 2-D Coding Improves the Reliability}
\begin{figure}[htbp]
  \centering
  \begin{tikzpicture}
    \definecolor{myblued}{RGB}{0,114,189}
    \definecolor{myred}{RGB}{217,83,25}
    \definecolor{myyellow}{RGB}{237,137,32}
    \definecolor{mypurple}{RGB}{126,47,142}
    \definecolor{myblues}{RGB}{77,190,238}
    \definecolor{mygreen}{RGB}{32,134,48}
    \definecolor{mypink}{RGB}{255,62,150}
      \pgfplotsset{
        label style = {font=\fontsize{9pt}{7.2}\selectfont},
        tick label style = {font=\fontsize{7pt}{7.2}\selectfont}
      }

    \begin{axis}[
        scale = 1,
        ymode=log,
        xmin=10.0,xmax=18.0,
        ymin=3.0E-06,ymax=0.8,
        xlabel={$E_b/N_0$ [dB]}, xlabel style={yshift=0.1em},
        ylabel={FER}, ylabel style={yshift=-0.75em},
        xtick={10,11,12,13,14,15,16,17,18},
        xticklabels={10,11,12,13,14,15,16,17,18},
        grid=both,
        ymajorgrids=true,
        xmajorgrids=true,
        grid style=dashed,
        width=0.9\linewidth,
        thick,
        mark size=1,
        legend style={
          nodes={scale=0.6, transform shape},
          anchor={center},
          cells={anchor=west},
          column sep= 1mm,
          row sep= -0.25mm,
          font=\fontsize{8pt}{7.2}\selectfont,
        },
        legend columns=1,
        legend pos= south west,
    ]

    \addplot[
        color=myyellow,
        mark=square*,
        densely dotted,
        every mark/.append style={solid},
        line width=0.25mm,
        mark size=1.9,
        fill opacity=0,
    ]
    table {
     10     0.699300699300699	
     11     0.436681222707424	
     12     0.251256281407035	
     13     0.174216027874564	
     14     0.106044538706257	    
     15     0.0493339911198816	     
     16     0.0279720279720280	    
     17     0.0159159637116027	    
     18     0.00657462195923734	    
     19     0.00308194902456313	    
     20     0.00114005586273727
    };
    \addlegendentry{Time-domain 1-D, $N_s=16$, $R_s=1/4$}

    \addplot[
        color=myblues,
        mark=diamond*,
        densely dotted,
        every mark/.append style={solid},
        line width=0.25mm,
        mark size=1.9,
        fill opacity=0,
    ]
    table {
     10     0.492610837438424	
     11     0.238663484486874	
     12     0.130548302872063	
     13     0.0652741514360313	
     14     0.0381533765738268	    
     15     0.0176772140710624	     
     16     0.00905305087814593	
     17     0.00473596968979399	    
     18     0.00241027742293138	    
     19     0.000792148226776194	
     20     0.000248765501200294
    };
    \addlegendentry{Time-domain 1-D, $N_s=32$, $R_s=1/4$}

    \addplot[
        color=myred,
        mark=triangle*,
        densely dotted,
        every mark/.append style={solid},
       line width=0.25mm,
        mark size=1.9,
        fill opacity=0,
    ]
    table {
     10     0.170068027210884	
     11     0.063011972274732	
     12     0.0226500566251416	
     13     0.00828706389326262	
     14     0.00348298561526941	    
     15     0.00138969954695795	     
     16     0.000373254103928873	
     17     0.000133505109908082   	
     18     3.40000000000000e-05	
     19     1.40000000000000e-05
    };
    \addlegendentry{Time-domain 1-D, $N_s=64$, $R_s=1/4$}

    
    \addplot[
        color=myyellow,
        mark=square*,
        line width=0.25mm,
        mark size=1.9,
        fill opacity=0,
    ]
    table {
     10     0.943396226415094	
     11     0.769230769230769	
     12     0.312500000000000	
     13     0.0811030008110300	
     14     0.0208420175072947	    
     15     0.00366985944438328	     
     16     0.000736800223987268	
     17     0.000162400671689178	
     18     2.70000000000000e-05	
     19     1.00000000000000e-06
    };
    \addlegendentry{2-D, $N^\mathrm{time}=16$,  $N^\mathrm{space}=64$, $R^\mathrm{time}=R^\mathrm{space}=1/2$}
    
    \addplot[
        color=myblues,
        mark=diamond*,
        line width=0.25mm,
        mark size=1.9,
        fill opacity=0,
    ]
    table {
     10     1                 
     11     0.653594771241830   
     12     0.288184438040346   
     13     0.0701754385964912  
     14     0.0156298843388559      
     15     0.00186511489107729      
     16     0.000381177457165183   
     17     5.30000000000000e-05    
     18     3.00000000000000e-06
    };
    \addlegendentry{2-D, $N^\mathrm{time}=32$,  $N^\mathrm{space}=64$, $R^\mathrm{time}=R^\mathrm{space}=1/2$}
    
    \addplot[
        color=myred,
        mark=triangle*,
       line width=0.25mm,
        mark size=1.9,
        fill opacity=0,
    ]
    table {
     10     0.909090909090909 
     11     0.719424460431655   
     12     0.292397660818713   
     13     0.0634115409004439  
     14     0.0110350915912602      
     15     0.00169782169476562      
     16     0.000293682306228414   
     17     4.40355660588535e-05    
     18     7.00000000000000e-06
    };
    \addlegendentry{2-D, $N^\mathrm{time}=64$,  $N^\mathrm{space}=64$, $R^\mathrm{time}=R^\mathrm{space}=1/2$}
    
    \end{axis}
    \end{tikzpicture}
  \caption{
  FER performance comparison between the time-domain 1-D coding and spatiotemporal 2-D coding with different time-domain code lengths.
  }\label{fig:FER_sameRate}
\end{figure}
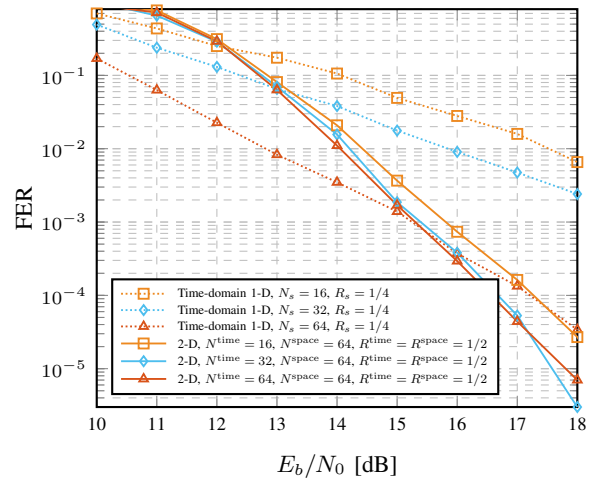

According to the results shown in Fig.~\ref{fig:FER_sameRate}, it is clear that the introduction of spatiotemporal 2-D coding will consistently bring reliability advantages for different time-domain code lengths. For comparison fairness, the spatiotemporal 2-D coding keeps the same time-domain code length ($N^{\text{time}}$) and overall code rate ($R^{\text{time}} \times R^{\text{space}}=R_s=1/4$) as its time-domain 1-D counterpart as in Fig.~\ref{fig:trellisA}. For instance, for a short length of $N^{\text{time}}=16$, the performance gain of 2-D coding is more than $3$\,dB at $\text{FER}=10^{-3}$.

\subsection{Space-Domain Coding Takes a Dominant Role}
Fig.~\ref{fig:FER_sameRate} offers us more information. Comparing error performance of spatiotemporal 2-D coding curves with different $N^{\text{time}}$, it is shown that the curve varies little while $N^{\text{time}}$ changing. The reason is that the space-domain coding takes a dominant role in spatiotemporal coding when $N^{\text{time}}$ is relatively small. As long as the space-domain coding stays the same, the performance stays similar, no matter how other parameters such as $N^{\text{time}}$ change.

This observation inspires us to think of approaches to further improve the reliability of the spatiotemporal 2-D coding. Two immediate approaches are 1) increasing the space-domain code length $N^{\text{space}}$, and 2) reducing the space-domain code rate $R^{\text{space}}$.

\subsection{Increasing $N^{\text{space}}$ Improves the Reliability}
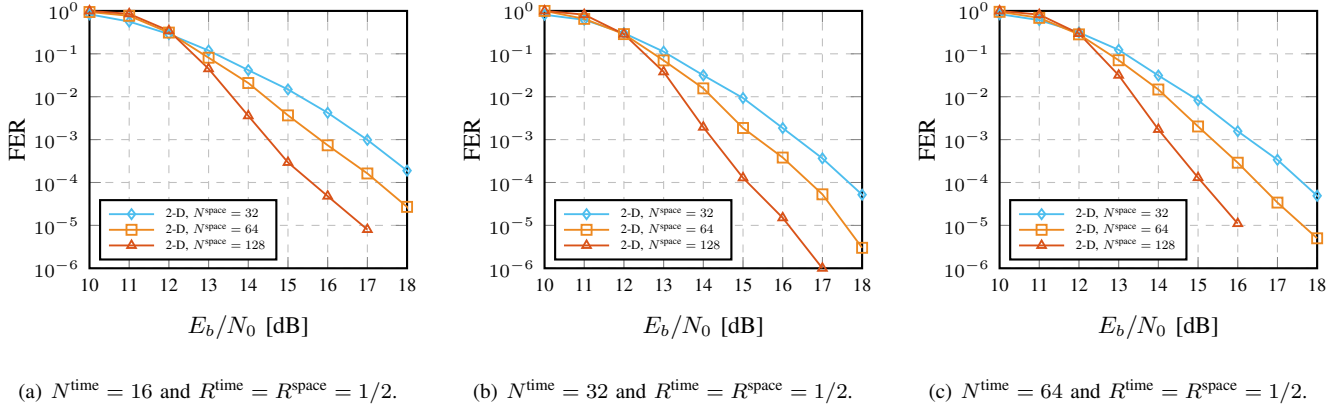
\begin{figure*}
    \centering
    \subfigure[$N^\text{time}=16$ and $R^\text{time}=R^\text{space}=1/2$.]{
    \begin{minipage}[t]{0.32\linewidth}
    \centering
    \begin{tikzpicture}
    \definecolor{myblued}{RGB}{0,114,189}
    \definecolor{myred}{RGB}{217,83,25}
    \definecolor{myyellow}{RGB}{237,137,32}
    \definecolor{mypurple}{RGB}{126,47,142}
    \definecolor{myblues}{RGB}{77,190,238}
    \definecolor{mygreen}{RGB}{32,134,48}
    \definecolor{mypink}{RGB}{255,62,150}
      \pgfplotsset{
        label style = {font=\fontsize{9pt}{7.2}\selectfont},
        tick label style = {font=\fontsize{7pt}{7.2}\selectfont}
      }

    \begin{axis}[
        scale = 1,
        ymode=log,
        xmin=10.0,xmax=18.0,
        ymin=1.0E-06,ymax=1,
        xlabel={$E_b/N_0$ [dB]}, xlabel style={yshift=0.1em},
        ylabel={FER}, ylabel style={yshift=-0.75em},
        xtick={10,11,12,13,14,15,16,17,18},
        xticklabels={10,11,12,13,14,15,16,17,18},
        ytick={1,0.1,0.01,0.001,0.0001,0.00001,0.000001},
        yticklabels={$10^{0}$,$10^{-1}$,$10^{-2}$,$10^{-3}$,$10^{-4}$,$10^{-5}$,$10^{-6}$},
        ymajorgrids=true,
        xmajorgrids=true,
        grid style=dashed,
        width=\linewidth,
        thick,
        mark size=1,
        legend style={
          nodes={scale=0.6, transform shape},
          anchor={center},
          cells={anchor=west},
          column sep= 1mm,
          row sep= -0.25mm,
          font=\fontsize{8pt}{7.2}\selectfont,
        },
        legend columns=1,
        legend pos= south west,
    ]

    \addplot[
        color=myblues,
        mark=diamond*,
        line width=0.25mm,
        mark size=1.9,
        fill opacity=0,
    ]
    table {
     10     0.833333333333333	
     11     0.564971751412429	
     12     0.289017341040462	
     13     0.119189511323004	
     14     0.0416493127863390	    
     15     0.0147623265426631	     
     16     0.00422743606002959	
     17     0.000984445757038787	
     18     0.000189134953463345	
     19     2.90000000000000e-05	
     20     2.00000000000000e-06
    };
    \addlegendentry{2-D, $N^\text{space}=32$}

    \addplot[
        color=myyellow,
        mark=square*,
        line width=0.25mm,
        mark size=1.9,
        fill opacity=0,
    ]
    table {
     10     0.943396226415094	
     11     0.769230769230769	
     12     0.312500000000000	
     13     0.0811030008110300	
     14     0.0208420175072947	    
     15     0.00366985944438328	     
     16     0.000736800223987268	
     17     0.000162400671689178	
     18     2.70000000000000e-05	
     19     1.00000000000000e-06
    };
    \addlegendentry{2-D, $N^\text{space}=64$}

    \addplot[
        color=myred,
        mark=triangle*,
       line width=0.25mm,
        mark size=1.9,
        fill opacity=0,
    ]
    table {
     10     1	                
     11     0.854700854700855	
     12     0.344827586206897	
     13     0.0448028673835125	
     14     0.00358757264834613	    
     15     0.000293329109394158	 
     16     4.80000000000000e-05	
     17     8.00000000000000e-06
    };
    \addlegendentry{2-D, $N^\text{space}=128$}
    
    \end{axis}
    \end{tikzpicture}
    \label{results_0301}
    \end{minipage}}
    \subfigure[$N^\text{time}=32$ and $R^\text{time}=R^\text{space}=1/2$.]{
    \begin{minipage}[t]{0.32\linewidth}
    \centering
    \begin{tikzpicture}
    \definecolor{myblued}{RGB}{0,114,189}
    \definecolor{myred}{RGB}{217,83,25}
    \definecolor{myyellow}{RGB}{237,137,32}
    \definecolor{mypurple}{RGB}{126,47,142}
    \definecolor{myblues}{RGB}{77,190,238}
    \definecolor{mygreen}{RGB}{32,134,48}
    \definecolor{mypink}{RGB}{255,62,150}
      \pgfplotsset{
        label style = {font=\fontsize{9pt}{7.2}\selectfont},
        tick label style = {font=\fontsize{7pt}{7.2}\selectfont}
      }

    \begin{axis}[
        scale = 1,
        ymode=log,
        xmin=10.0,xmax=18.0,
        ymin=1.0E-06,ymax=1,
        xlabel={$E_b/N_0$ [dB]}, xlabel style={yshift=0.1em},
        ylabel={FER}, ylabel style={yshift=-0.75em},
        xtick={10,11,12,13,14,15,16,17,18},
        xticklabels={10,11,12,13,14,15,16,17,18},
        ytick={1,0.1,0.01,0.001,0.0001,0.00001,0.000001},
        yticklabels={$10^{0}$,$10^{-1}$,$10^{-2}$,$10^{-3}$,$10^{-4}$,$10^{-5}$,$10^{-6}$},
        grid=both,
        ymajorgrids=true,
        xmajorgrids=true,
        grid style=dashed,
        width=\linewidth,
        thick,
        mark size=1,
        legend style={
          nodes={scale=0.6, transform shape},
          anchor={center},
          cells={anchor=west},
          column sep= 1mm,
          row sep= -0.25mm,
          font=\fontsize{8pt}{7.2}\selectfont,
        },
        legend columns=1,
        legend pos= south west,
    ]

    \addplot[
        color=myblues,
        mark=diamond*,
        line width=0.25mm,
        mark size=1.9,
        fill opacity=0,
    ]
    table {
     10     0.813008130081301	
     11     0.621118012422360	
     12     0.303951367781155	
     13     0.112233445566779	
     14     0.0315059861373661	    
     15     0.00930319099451112   	 
     16     0.00185673437558023	
     17     0.000364803735590252	
     18     5.20000000000000e-05	
     19     7.00000000000000e-06	
     20     2.00000000000000e-06
    };
    \addlegendentry{2-D, $N^\text{space}=32$}

    \addplot[
        color=myyellow,
        mark=square*,
        line width=0.25mm,
        mark size=1.9,
        fill opacity=0,
    ]
    table {
     10     1                 
     11     0.653594771241830   
     12     0.288184438040346   
     13     0.0701754385964912  
     14     0.0156298843388559      
     15     0.00186511489107729      
     16     0.000381177457165183   
     17     5.30000000000000e-05    
     18     3.00000000000000e-06
    };
    \addlegendentry{2-D, $N^\text{space}=64$}

    \addplot[
        color=myred,
        mark=triangle*,
       line width=0.25mm,
        mark size=1.9,
        fill opacity=0,
    ]
    table {
     10     1	                
     11     0.819672131147541	
     12     0.287356321839080	
     13     0.0379506641366224	
     14     0.00194363459669582	    
     15     0.000128220416023962	 
     16     1.50000000000000e-05	
     17     1.00000000000000e-06
    };
    \addlegendentry{2-D, $N^\text{space}=128$}
    
    \end{axis}
    \end{tikzpicture}
    \label{results_0302}
    \end{minipage}}
    \subfigure[$N^\text{time}=64$ and $R^\text{time}=R^\text{space}=1/2$.]{
    \begin{minipage}[t]{0.32\linewidth}
    \centering
    \begin{tikzpicture}
    \definecolor{myblued}{RGB}{0,114,189}
    \definecolor{myred}{RGB}{217,83,25}
    \definecolor{myyellow}{RGB}{237,137,32}
    \definecolor{mypurple}{RGB}{126,47,142}
    \definecolor{myblues}{RGB}{77,190,238}
    \definecolor{mygreen}{RGB}{32,134,48}
    \definecolor{mypink}{RGB}{255,62,150}
      \pgfplotsset{
        label style = {font=\fontsize{9pt}{7.2}\selectfont},
        tick label style = {font=\fontsize{7pt}{7.2}\selectfont}
      }

    \begin{axis}[
        scale = 1,
        ymode=log,
        xmin=10.0,xmax=18.0,
        ymin=1.0E-06,ymax=1,
        xlabel={$E_b/N_0$ [dB]}, xlabel style={yshift=0.1em},
        ylabel={FER}, ylabel style={yshift=-0.75em},
        xtick={10,11,12,13,14,15,16,17,18},
        xticklabels={10,11,12,13,14,15,16,17,18},
        ytick={1,0.1,0.01,0.001,0.0001,0.00001,0.000001},
        yticklabels={$10^{0}$,$10^{-1}$,$10^{-2}$,$10^{-3}$,$10^{-4}$,$10^{-5}$,$10^{-6}$},
        grid=both,
        ymajorgrids=true,
        xmajorgrids=true,
        grid style=dashed,
        width=\linewidth,
        thick,
        mark size=1,
        legend style={
          nodes={scale=0.6, transform shape},
          anchor={center},
          cells={anchor=west},
          column sep= 1mm,
          row sep= -0.25mm,
          font=\fontsize{8pt}{7.2}\selectfont,
        },
        legend columns=1,
        legend pos= south west,
        ]

    \addplot[
        color=myblues,
        mark=diamond*,
        line width=0.25mm,
        mark size=1.9,
        fill opacity=0,
    ]
    table {
     10     0.840336134453782	
     11     0.609756097560976	
     12     0.314465408805031	
     13     0.123762376237624	
     14     0.0312012480499220	    
     15     0.00826309700875888	     
     16     0.00156759468271884	
     17     0.000338055975308392	
     18     4.90000000000000e-05	
     19     1.00000000000000e-05	
     20     1.00000000000000e-06
    };
    \addlegendentry{2-D, $N^\text{space}=32$}

    \addplot[
        color=myyellow,
        mark=square*,
        line width=0.25mm,
        mark size=1.9,
        fill opacity=0,
    ]
    table {
     10     0.952380952380952	
     11     0.689655172413793	
     12     0.284090909090909	
     13     0.0708717221828490	
     14     0.0147145379635079	    
     15     0.00202334945268397	     
     16     0.000289024023676848	
     17     3.40000000000000e-05	
     18     5.00000000000000e-06
    };
    \addlegendentry{2-D, $N^\text{space}=64$}

    \addplot[
        color=myred,
        mark=triangle*,
       line width=0.25mm,
        mark size=1.9,
        fill opacity=0,
    ]
    table {
     10     1	                
     11     0.819672131147541	
     12     0.296735905044510	
     13     0.0316555872111428	
     14     0.00172500043125011	    
     15     0.000129759918199348	 
     16     1.10000000000000e-05
    };
    \addlegendentry{2-D, $N^\text{space}=128$}
    
    \end{axis}
    \end{tikzpicture}
    \label{results_0303}
    \end{minipage}}
	\caption{
	FER performance comparison of the spatiotemporal 2-D coding with different space-domain code lengths.
	} \label{fig:reults03}
\end{figure*}

According to the results in Fig.~\ref{fig:reults03}, it can be seen that for the same $N^{\text{time}}$, higher $N^{\text{space}}$ brings higher reliability. This holds for different values of $N^{\text{time}}$. It is noted that here, in the three plots $N_t=N_r/2=N^{\text{space}}$, and $N_t \times N_r = 64 \times 128$ does not holds here. Therefore, for applications requiring higher reliability, designers can simply increase the number of antennas for space-domain coding.

\subsection{Decreasing $R^{\text{space}}$ Improves the Reliability}
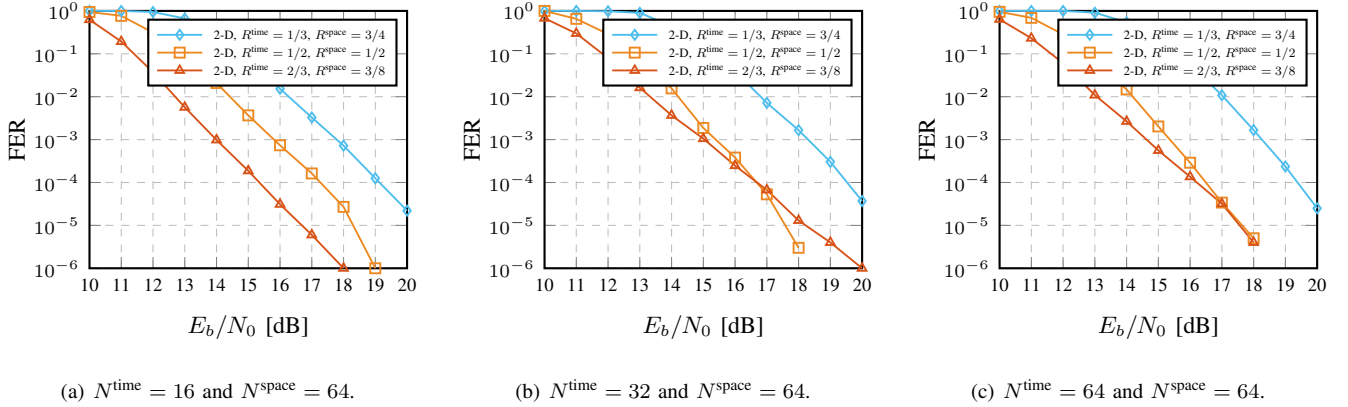
\begin{figure*}
    \centering
    \subfigure[$N^\text{time}=16$ and $N^\text{space}=64$.]{
    \begin{minipage}[t]{0.32\linewidth}
    \centering
    \begin{tikzpicture}
    \definecolor{myblued}{RGB}{0,114,189}
    \definecolor{myred}{RGB}{217,83,25}
    \definecolor{myyellow}{RGB}{237,137,32}
    \definecolor{mypurple}{RGB}{126,47,142}
    \definecolor{myblues}{RGB}{77,190,238}
    \definecolor{mygreen}{RGB}{32,134,48}
    \definecolor{mypink}{RGB}{255,62,150}
      \pgfplotsset{
        label style = {font=\fontsize{9pt}{7.2}\selectfont},
        tick label style = {font=\fontsize{7pt}{7.2}\selectfont}
      }

    \begin{axis}[
        scale = 1,
        ymode=log,
        xmin=10.0,xmax=20.0,
        ymin=1.0E-06,ymax=1,
        xlabel={$E_b/N_0$ [dB]}, xlabel style={yshift=0.1em},
        ylabel={FER}, ylabel style={yshift=-0.75em},
        xtick={10,11,12,13,14,15,16,17,18,19,20},
        xticklabels={10,11,12,13,14,15,16,17,18,19,20},
        ytick={1,0.1,0.01,0.001,0.0001,0.00001,0.000001},
        yticklabels={$10^{0}$,$10^{-1}$,$10^{-2}$,$10^{-3}$,$10^{-4}$,$10^{-5}$,$10^{-6}$},
        ymajorgrids=true,
        xmajorgrids=true,
        grid style=dashed,
        width=\linewidth,
        thick,
        mark size=1,
        legend style={
          nodes={scale=0.6, transform shape},
          anchor={center},
          cells={anchor=west},
          column sep= 1mm,
          row sep= -0.25mm,
          font=\fontsize{8pt}{7.2}\selectfont,
        },
        legend columns=1,
        legend pos= north east,
    ]

    \addplot[
        color=myblues,
        mark=diamond*,
        line width=0.25mm,
        mark size=1.9,
        fill opacity=0,
    ]
    table {
     10     1	                
     11     1	                
     12     0.943396226415094	
     13     0.666666666666667	
     14     0.278551532033426	    
     15     0.0740192450037010	     
     16     0.0152835090936879	    
     17     0.00327922610263978	    
     18     0.000722611227933260   
     19     0.000125168038091137	  
     20     2.20000000000000e-05
    };
    \addlegendentry{2-D, $R^\text{time}=1/3$, $R^\text{space}=3/4$}

    \addplot[
        color=myyellow,
        mark=square*,
        line width=0.25mm,
        mark size=1.9,
        fill opacity=0,
    ]
    table {
     10     0.943396226415094	
     11     0.769230769230769	
     12     0.312500000000000	
     13     0.0811030008110300	
     14     0.0208420175072947	    
     15     0.00366985944438328	     
     16     0.000736800223987268	
     17     0.000162400671689178	
     18     2.70000000000000e-05	
     19     1.00000000000000e-06
    };
    \addlegendentry{2-D, $R^\text{time}=1/2$, $R^\text{space}=1/2$}

    \addplot[
        color=myred,
        mark=triangle*,
       line width=0.25mm,
        mark size=1.9,
        fill opacity=0,
    ]
    table {
     10     0.625000000000000	
     11     0.193423597678917	
     12     0.0372856077554064	
     13     0.00566829157691872	
     14     0.000981575821824357	
     15     0.000186832424393589	 
     16     3.10000000000000e-05	
     17     6.00000000000000e-06	
     18     1.00000000000000e-06
    };
    \addlegendentry{2-D, $R^\text{time}=2/3$, $R^\text{space}=3/8$}
    
    \end{axis}
    \end{tikzpicture}
    \label{results_0301}
    \end{minipage}}
    \subfigure[$N^\text{time}=32$ and $N^\text{space}=64$.]{
    \begin{minipage}[t]{0.32\linewidth}
    \centering
    \begin{tikzpicture}
    \definecolor{myblued}{RGB}{0,114,189}
    \definecolor{myred}{RGB}{217,83,25}
    \definecolor{myyellow}{RGB}{237,137,32}
    \definecolor{mypurple}{RGB}{126,47,142}
    \definecolor{myblues}{RGB}{77,190,238}
    \definecolor{mygreen}{RGB}{32,134,48}
    \definecolor{mypink}{RGB}{255,62,150}
      \pgfplotsset{
        label style = {font=\fontsize{9pt}{7.2}\selectfont},
        tick label style = {font=\fontsize{7pt}{7.2}\selectfont}
      }

    \begin{axis}[
        scale = 1,
        ymode=log,
        xmin=10.0,xmax=20.0,
        ymin=1.0E-06,ymax=1,
        xlabel={$E_b/N_0$ [dB]}, xlabel style={yshift=0.1em},
        ylabel={FER}, ylabel style={yshift=-0.75em},
        xtick={10,11,12,13,14,15,16,17,18,19,20},
        xticklabels={10,11,12,13,14,15,16,17,18,19,20},
        ytick={1,0.1,0.01,0.001,0.0001,0.00001,0.000001},
        yticklabels={$10^{0}$,$10^{-1}$,$10^{-2}$,$10^{-3}$,$10^{-4}$,$10^{-5}$,$10^{-6}$},
        ymajorgrids=true,
        xmajorgrids=true,
        grid style=dashed,
        width=\linewidth,
        thick,
        mark size=1,
        legend style={
          nodes={scale=0.6, transform shape},
          anchor={center},
          cells={anchor=west},
          column sep= 1mm,
          row sep= -0.25mm,
          font=\fontsize{8pt}{7.2}\selectfont,
        },
        legend columns=1,
        legend pos= north east,
    ]

    \addplot[
        color=myblues,
        mark=diamond*,
        line width=0.25mm,
        mark size=1.9,
        fill opacity=0,
    ]
    table {
     10     1	                
     11     1	                
     12     0.980392156862745	
     13     0.900900900900901	
     14     0.444444444444444	    
     15     0.135501355013550	     
     16     0.0391849529780564	    
     17     0.00717823558969205	    
     18     0.00165980613464347	    
     19     0.000302833002740639	
     20     3.70000000000000e-05
    };
    \addlegendentry{2-D, $R^\text{time}=1/3$, $R^\text{space}=3/4$}

    \addplot[
        color=myyellow,
        mark=square*,
        line width=0.25mm,
        mark size=1.9,
        fill opacity=0,
    ]
    table {
     10     1                 
     11     0.653594771241830   
     12     0.288184438040346   
     13     0.0701754385964912  
     14     0.0156298843388559      
     15     0.00186511489107729      
     16     0.000381177457165183   
     17     5.30000000000000e-05    
     18     3.00000000000000e-06
    };
    \addlegendentry{2-D, $R^\text{time}=1/2$, $R^\text{space}=1/2$}

    \addplot[
        color=myred,
        mark=triangle*,
       line width=0.25mm,
        mark size=1.9,
        fill opacity=0,
    ]
    table {
     10     0.675675675675676	
     11     0.298507462686567	
     12     0.0793021411578113	
     13     0.0161969549724652	
     14     0.00367836386375340	
     15     0.00105188971988177	
     16     0.000246900168385915	
     17     6.70000000000000e-05	
     18     1.30000000000000e-05	
     19     4.00000000000000e-06	
     20     1.00000000000000e-06
    };
    \addlegendentry{2-D, $R^\text{time}=2/3$, $R^\text{space}=3/8$}
    
    \end{axis}
    \end{tikzpicture}
    \label{results_0302}
    \end{minipage}}
    \subfigure[$N^\text{time}=64$ and $N^\text{space}=64$.]{
    \begin{minipage}[t]{0.32\linewidth}
    \centering
    \begin{tikzpicture}
    \definecolor{myblued}{RGB}{0,114,189}
    \definecolor{myred}{RGB}{217,83,25}
    \definecolor{myyellow}{RGB}{237,137,32}
    \definecolor{mypurple}{RGB}{126,47,142}
    \definecolor{myblues}{RGB}{77,190,238}
    \definecolor{mygreen}{RGB}{32,134,48}
    \definecolor{mypink}{RGB}{255,62,150}
      \pgfplotsset{
        label style = {font=\fontsize{9pt}{7.2}\selectfont},
        tick label style = {font=\fontsize{7pt}{7.2}\selectfont}
      }

    \begin{axis}[
        scale = 1,
        ymode=log,
        xmin=10.0,xmax=20.0,
        ymin=1.0E-06,ymax=1,
        xlabel={$E_b/N_0$ [dB]}, xlabel style={yshift=0.1em},
        ylabel={FER}, ylabel style={yshift=-0.75em},
        xtick={10,11,12,13,14,15,16,17,18,19,20},
        xticklabels={10,11,12,13,14,15,16,17,18,19,20},
        ytick={1,0.1,0.01,0.001,0.0001,0.00001,0.000001},
        yticklabels={$10^{0}$,$10^{-1}$,$10^{-2}$,$10^{-3}$,$10^{-4}$,$10^{-5}$,$10^{-6}$},
        ymajorgrids=true,
        xmajorgrids=true,
        grid style=dashed,
        width=\linewidth,
        thick,
        mark size=1,
        legend style={
          nodes={scale=0.6, transform shape},
          anchor={center},
          cells={anchor=west},
          column sep= 1mm,
          row sep= -0.25mm,
          font=\fontsize{8pt}{7.2}\selectfont,
        },
        legend columns=1,
        legend pos= north east,
    ]

    \addplot[
        color=myblues,
        mark=diamond*,
        line width=0.25mm,
        mark size=1.9,
        fill opacity=0,
    ]
    table {
     10     1	                
     11     1	                
     12     1	                
     13     0.909090909090909	
     14     0.543478260869565	    
     15     0.229357798165138	     
     16     0.0556792873051225	    
     17     0.0108061378863194	    
     18     0.00166336765415260	    
     19     0.000236188856609745	
     20     2.50000000000000e-05
    };
    \addlegendentry{2-D, $R^\text{time}=1/3$, $R^\text{space}=3/4$}

    \addplot[
        color=myyellow,
        mark=square*,
        line width=0.25mm,
        mark size=1.9,
        fill opacity=0,
    ]
    table {
     10     0.952380952380952	
     11     0.689655172413793	
     12     0.284090909090909	
     13     0.0708717221828490	
     14     0.0147145379635079	    
     15     0.00202334945268397	     
     16     0.000289024023676848	
     17     3.40000000000000e-05	
     18     5.00000000000000e-06
    };
    \addlegendentry{2-D, $R^\text{time}=1/2$, $R^\text{space}=1/2$}

    \addplot[
        color=myred,
        mark=triangle*,
       line width=0.25mm,
        mark size=1.9,
        fill opacity=0,
    ]
    table {
     10     0.636942675159236	
     11     0.230946882217090	
     12     0.0605693519079346	
     13     0.0108932461873638	
     14     0.00266283218831549	    
     15     0.000558634250983196	 
     16     0.000135834868267345	
     17     3.20000000000000e-05	
     18     4.00000000000000e-06	
    };
    \addlegendentry{2-D, $R^\text{time}=2/3$, $R^\text{space}=3/8$}
    
    \end{axis}
    \end{tikzpicture}
    \label{results_0303}
    \end{minipage}}
	\caption{
	FER performance comparison of the spatiotemporal 2-D coding with different space-domain code rates.
	} \label{fig:reults04}
\end{figure*}

According to the results in Fig.~\ref{fig:reults04}, it can be seen that for the same $N^{\text{time}}$ and $N^{\text{space}}$, lower $R^{\text{space}}$ brings higher reliability. This holds for different values of $N^{\text{time}}$. It is noted that, for fair comparison, the time-domain code rate $R^{\text{time}}$ increases while $R^{\text{space}}$ decreasing, to make sure $R^{\text{time}} \times R^{\text{space}}=1/4$ is a constant. Therefore, for applications requiring higher reliability, designers can also decrease the code rate of space-domain coding.

\subsection{Commutativity of Time/Space-Domain Coding}
\begin{figure}[htbp]
  \centering
  \begin{tikzpicture}
    \definecolor{myblued}{RGB}{0,114,189}
    \definecolor{myred}{RGB}{217,83,25}
    \definecolor{myyellow}{RGB}{237,137,32}
    \definecolor{mypurple}{RGB}{126,47,142}
    \definecolor{myblues}{RGB}{77,190,238}
    \definecolor{mygreen}{RGB}{32,134,48}
    \definecolor{mypink}{RGB}{255,62,150}
      \pgfplotsset{
        label style = {font=\fontsize{9pt}{7.2}\selectfont},
        tick label style = {font=\fontsize{7pt}{7.2}\selectfont}
      }

    \begin{axis}[
        scale = 1,
        ymode=log,
        xmin=10.0,xmax=19.0,
        ymin=3.0E-06,ymax=1,
        xlabel={$E_b/N_0$ [dB]}, xlabel style={yshift=0.1em},
        ylabel={FER}, ylabel style={yshift=-0.75em},
        xtick={10,11,12,13,14,15,16,17,18,19},
        xticklabels={10,11,12,13,14,15,16,17,18,19},
        grid=both,
        ymajorgrids=true,
        xmajorgrids=true,
        grid style=dashed,
        width=0.9\linewidth,
        thick,
        mark size=1,
        legend style={
          nodes={scale=0.6, transform shape},
          anchor={center},
          cells={anchor=west},
          column sep= 1mm,
          row sep= -0.25mm,
          font=\fontsize{8pt}{7.2}\selectfont,
        },
        legend columns=1,
        legend pos= south west,
    ]

    \addplot[
        color=myyellow,
        mark=square*,
        densely dotted,
        every mark/.append style={solid},
        line width=0.25mm,
        mark size=1.9,
        fill opacity=0,
    ]
    table {
     10     0.699300699300699	
     11     0.436681222707424	
     12     0.251256281407035	
     13     0.174216027874564	
     14     0.106044538706257	    
     15     0.0493339911198816	     
     16     0.0279720279720280	    
     17     0.0159159637116027	    
     18     0.00657462195923734	    
     19     0.00308194902456313	    
     20     0.00114005586273727
    };
    \addlegendentry{Time-domain 1-D, $N_s=16$, $R_s=1/4$}

    \addplot[
        color=myyellow,
        mark=square*,
        line width=0.25mm,
        mark size=1.9,
        fill opacity=0,
    ]
    table {
     10     0.943396226415094	
     11     0.769230769230769	
     12     0.312500000000000	
     13     0.0811030008110300	
     14     0.0208420175072947	    
     15     0.00366985944438328	     
     16     0.000736800223987268	
     17     0.000162400671689178	
     18     2.70000000000000e-05	
     19     1.00000000000000e-06
    };
    \addlegendentry{2-D, T-S, $N^\text{time}=16$,  $N^\text{space}=64$, $R^\text{time}=R^\text{space}=1/2$}
    
    \addplot[
        color=myblues,
        mark=diamond*,
        line width=0.25mm,
        mark size=1.9,
        fill opacity=0,
    ]
    table {
     10     0.740740740740741	
     11     0.400000000000000	
     12     0.139664804469274	
     13     0.0448833034111311	
     14     0.0160823415889354	    
     15     0.00434593654932638	     
     16     0.00116691560866318	
     17     0.000312292130550602	
     18     7.00000000000000e-05	
     19     1.20000000000000e-05
    };
    \addlegendentry{2-D, S-T, $N^\text{space}=64$, $N^\text{time}=16$, $R^\text{time}=R^\text{space}=1/2$}
    
    \end{axis}
    \end{tikzpicture}
  \caption{
  FER performance comparison between the time-space (T-S) mode and the space-time (S-T) mode.
  }\label{fig:results05}
\end{figure}
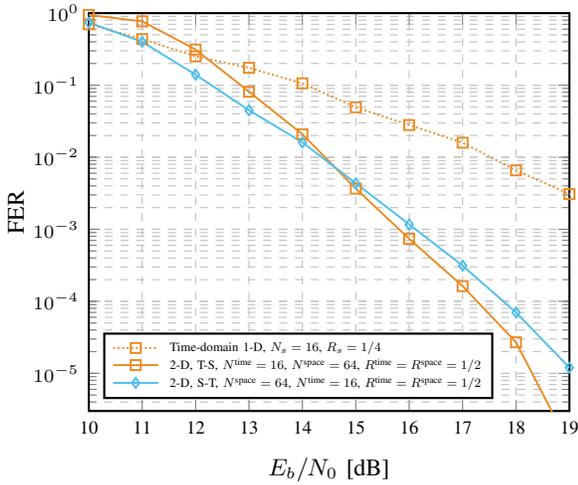

According to Section \ref{subsec:flexible2DTS}, it is known that the time-space mode and space-time mode are differentiated in the encoding order. Therefore, their performance will be similar, which is verified by Fig.~\ref{fig:results05}. In real applications, users can decide to choose which one according to the specific requirements.

	\section{Discussions}\label{sec:discussions}

In this section, some discussions are provided to offer useful insights for readers.

\subsubsection{Flexibility}
The application of the proposed coding scheme is not limited to the very low latency reliable MIMO transmission. This scheme can flexibly trade-off between latency and reliability by configuring the code rate, code length, and parameters $S$ and $T$. For example, if $S=1$ and $R^{\text{space}}=1$, the proposed spatiotemporal 2-D coding will become time domain 1-D coding, which is suitable for the eMBB case. In other words, the proposed scheme offers a uniform scheme which can meet the requirements of vast applications including both eMBB and URLLC cases.

\subsubsection{Versatility}
Though polar codes are employed as a running example for better understanding of the proposed scheme, the channel codes involved do not necessarily be polar codes and can be freely chosen. Second, the proposed scheme only focuses on the channel coding part of baseband processing, therefore can be directly used by any MIMO systems including 5G, 5G Evolution, 6G, and so on, without changing the other parts of the existing systems.

\subsubsection{Theoretical Basis}
The proposed coding scheme can be treated as the application of product codes and their variants in massive MIMO systems. Therefore, the proposed scheme can further vary based on similar codes such as interleaved codes, concatenated product codes, and so on. The existing results of coding theory can be used to further improve the proposed spatiotemporal 2-D coding.

\subsubsection{Implementations}
Thanks to the regular structure of the 2-D codeword trellis, the proposed coding scheme is implementation friendly. Still there are implementation issues required further research. One issue is the implementation of iterative decoder of the spatiotemporal 2-D codes. Another issue is how the choice of block partition of codeword trellis and specific codes will affect the implementation efficiency. For example, should the channel codes of both dimensions be the same or not? If not, how to improve the hardware utilization by module sharing? 

	\section{Conclusions}\label{sec:conclusion}

In this paper, a spatiotemporal 2-D channel coding is first proposed for very low latency reliable MIMO transmission. Numerical simulations have demonstrated its advantages of performance, latency, and flexibility over the state-of-the-art schemes. With good versatility, the proposed coding scheme has wide application prospect. Future work will be directed to the hardware implementations and systematic applications. 

	\section{Acknowledgement}\label{sec:ack}
	The authors thank Huizheng Wang, Houren Ji, Xu Pang, and Leyu Zhang for their help in composing this paper.


	\bibliographystyle{IEEEtran}
	\bibliography{./bib/IEEEabrv,./bib/IEEEBib}
	
\end{document}